\documentclass[floatfix,twocolumn,aps,showpacs,superscriptaddress,longbibliography,10]{revtex4-1}

\usepackage{graphicx}
\usepackage{multirow,booktabs}
\usepackage{amsmath,amssymb,latexsym,MnSymbol}
\usepackage[colorlinks=true, urlcolor=black,linkcolor=black,citecolor=blue]{hyperref}

\makeatletter
\let\cat@comma@active\@empty
\makeatother

\usepackage{lineno}
\usepackage{verbatim}
\begin{document}

\title{The role of city size and urban metrics on crime modeling}

\author{Luiz G. A. Alves}\email{lgaalves@usp.br}
\affiliation{Institute of Mathematics and Computer Science, University of S\~ao Paulo, S\~ao Carlos, SP 13566-590, Brazil}

\author{Haroldo V. Ribeiro}
\affiliation{Departamento de F\'isica, Universidade Estadual de Maring\'a, Maring\'a, PR, 87020-900, Brazil}
\author{Francisco A. Rodrigues}
\affiliation{Institute of Mathematics and Computer Science, University of S\~ao Paulo, S\~ao Carlos, SP 13566-590, Brazil}

\date{\today}

\begin{abstract}
Unveiling the relationships between crime and socioeconomic factors is crucial for modeling and preventing these illegal activities. Recently, a significant advance has been made in understanding the influence of urban metrics on the levels of crime in different urban systems. In this chapter, we show how the dynamics of crime growth rate and the number of crime in cities are related to cities' size. We also discuss the role of urban metrics in crime modeling within the framework of the urban scaling hypothesis, where a data-driven approach is proposed for modeling crime. This model provides several insights into the mechanism ruling the dynamics of crime and can assist policymakers in making better decisions on resource allocation and help crime prevention.
\end{abstract}

\maketitle

\tableofcontents
\section{Introduction}
\label{sec:0}
Crime activities cause large economic losses for cities, companies, and individuals. People's well-being in cities can dramatically decrease with the increasing feeling of insecurity caused by criminal actions that happen in areas close to where they live~\cite{OECD2016wellbeing}. Actually, a well-known pattern of criminal activity called ``broken windows theory''~\cite{Wilson1982Broken}, suggest that degraded urban environments enhance criminal activities in their neighborhoods. In fact, empirical results indicate that criminals often commit new crimes in previously-visited places~\cite{Short2008Statistical}, and neighboring cities have correlated crime rates~\cite{Alves2015spatial}. 

There are multiple factors that may enhance criminal activities in certain geographical areas. Besides the spatial-driven forces governing crime, there are several attempts to model crime as a function of punishment, income, social inequality, gender, and other social variables~\cite{Gordon2010random}. Indeed, understanding the relationships between crime and socioeconomic indicators and how crime affects society organization is crucial for predicting and preventing these illegal activities. One of the first attempts to relate crime to socioeconomic metrics was made by Becker~\cite{becker1968crime} in 1968. Considering a social loss function due to the practice of illegal activities, he proposed that there is a probability of punishment per offense that depends on the frequency of action and gravity of the crime. In addition, there is also a cost for society for surveillance, apprehension, and punishment. He proposed an economic approach to evaluating the fraction of crimes that could be left unpunished to reduce the social costs of criminal punishment. 

Cities play an important role in defining the organization and shape of the interactions in our society. In fact, the bigger the city, the more it is capable of creating wealth and innovation. However, problems such as pollution, diseases, and crime also increase with the city size~\cite{Bettencourt2007PNAS,Alves2013distance,Oliveira2017scaling}. In this context, population size has an important role in defining the number of crimes expected in a city, but also other urban metrics can influence it, as we discuss in Sect.~\ref{sec:2}. 

One of the most interesting and recent findings about cities is related to scaling laws of urban metrics with the population size. Metrics such as crimes~\cite{Alves2013scaling,Alves2013distance}, GDP~\cite{Alves2013distance,Alves2014empirical}, illiteracy rates~\cite{Alves2015scaledmetric}, number of gas stations~\cite{Bettencourt2007PNAS}, and length of electrical cables~\cite{Bettencourt2007PNAS} scale with population size as a power-law function, $Y\approx N^\beta$, where $Y$ is the urban metric, $N$ is the population size, and $\beta$ is the scaling exponent. These nonlinearities are often completely ignored when trying to model a particular crime type. Because most of the urban indicators do not scale linearly with population size, models that do not take into account the ``natural'' scale of urban metrics are very likely to yield predictions and comparisons biased towards small cities for $\beta <1$, and towards bigger cities for $\beta >1$. 

Another effect of cities' size on crime is related to the growth rates of homicides~\cite{Alves2013scaling}. Considering the logarithmic growth rates, the variance of the growth rates can be described as a power-law decay function of the city size. This happens because big cities have more defined growth rates, whereas cities with small population size, only a single crime can produce a  large variation in the growth rate from one year to another. Thus, the variance of the growth rate of crime is expected to be larger in smaller cities. This effect of size and variance was also observed in the growth of several complex organizations, from firms~\cite{Stanley1998growth} and religious activity~\cite{Picoli2008universal} to paper's citations~\cite{Picoli2006scaling} and metabolic rates in biology~\cite{Labra2007scaling}.

In the next sections, we discuss the effects of population size on crime and the role of urban indicators in predicting these illegal activities. The chapter is divided into three sections. In the first one (Sect.~\ref{sec:1}), we discuss the scaling laws of crime and urban metrics with population size. Specifically, we study the scaling laws in the growth rates of crimes in cities and the allometric laws between urban metrics, including crime and population size. In the second one (Sect.~\ref{sec:2}), an alternative approach to incorporate the nonlinear effects of population size on crime modeling is presented. In particular, we describe a scale-adjusted metric that properly accounts for these nonlinearities. By using these scale-adjusted metrics, we propose a model to quantify the role of urban indicators in crime modeling. Finally, in Section~\ref{sec:3} we discuss some perspectives about crime modeling in the context of complex systems. 

\section{Effects of cities' size on crime and urban metrics}

Cities are a remarkable fingerprint of humans organization and interaction. In 2007, for the first time more than a half of the world's population was living in cities and by 2050, the United Nations~\cite{UN} estimates that two out of three people will be living in urban areas. As cities' size increase, problems such as crime~\cite{Alves2013distance,Alves2015scaledmetric}, diseases~\cite{Bettencourt2007PNAS,Fernando2017Spatial}, emissions of CO2~\cite{Oliveira2014CO2} scales in super-linear fashion with population size; whereas, metrics like illiteracy~\cite{Alves2013distance,Alves2015scaledmetric}, number of gas stations and length of power cables~\cite{Bettencourt2007PNAS} have a sub-liner relationship with cities size. Metrics related to individual needs such as sanitation~\cite{Alves2013distance,Alves2015scaledmetric}, electrical consumption and housing~\cite{Bettencourt2007PNAS} scales linearly with population size. Despite cities' size play an important role in urban metrics and crime, it was only recently that these non-linearities were introduced in the crime modeling~\cite{Bettencourt2007PNAS,Alves2013scaling,Alves2013distance,Alves2015scaledmetric}. 

One of the interesting patterns about the relationship of crime with city's size is related to the homicide growth rates. The crime growth rates (logarithmic returns) behave similarly to complex organizations, where the interactions between subsystems can be modeled and analyzed in terms of scaling laws, similar to physical systems where inanimate particles interacting to each other exhibits an emerging complex behavior~\cite{Alves2013scaling}. In Sub-sect.~\ref{subsec:1.1}, we discuss this relationship between crime and city' size and how it can be viewed in the framework of complex organizations of interacting subsystems. Next, in Sub-sect.~\ref{subsec:1.2}, we discuss the scaling laws between population size and urban metrics, with a special focus on modeling crime in the context where cities are considered non-extensive complex systems.

\label{sec:1}
\subsection{The dynamics of crime growth rates and its relation with population size}
\label{subsec:1.1}

Predicting and modeling crime growth in cities is crucial for preventing and creating better policies against illegal activities. A possible way of investigating the changes in crime is by considering the differences between the number of crimes from one year to another one. Let us define $H_i(t)$ as the number of crimes in the year $t$ at city $i$. We can analyze the changes rates of crime by considering the successive differences on the number of a particular type of crime, 
\begin{eqnarray}
Z_i(t)=H_i(t+\Delta t)- H_i(t),
\end{eqnarray}
where $\Delta t$ is the time interval between events. This approach does not require nonlinear or stochastic transformations, but it is seriously affected by the scale used when defining $H_i(t)$~\cite{Mantegna2000Econophysics}. For instance, it is common to find crime reports based on per capita measures or in terms of crime rates (e.g., crime per hundreds of inhabitants), which directly affects the rate of changes of crime in cities with different population size. To handle such scaling problem, we could consider the returns,  
\begin{eqnarray}
R_i(t)=\frac{H_i(t+\Delta t)- H_i(t)}{H_i(t)}=\frac{Z_i(t)}{H_i(t)}.
\end{eqnarray}
This approach properly accounts for losses and gains from one year to another but it is very sensitive to long-term changes. For instance, as the population grows, the number of crimes also increases which could directly affect the returns $R_i(t)$ as we increase $\Delta t$.

A common and effective approach for overcoming the above-exposed problems is by considering the logarithmic returns when dealing with time series of complex systems. This quantity is simply defined in terms of the logarithmic differences from one year to another, that is, 
\begin{eqnarray}\label{logreturn}
S_i(t)=\log \left[ \frac{H_i(t+\Delta t)}{H_i(t)} \right]. 
\end{eqnarray}
The advantage of this approach is that the average changes of scales are already incorporated in the definition of $S_i(t)$. On the other hand, according to Mantegna and Stanley~\cite{Mantegna2000Econophysics}, the problem of this quantity is that the correction of scale change would be correct only if the growth rate is constant. Usually, the growth of complex organizations fluctuates and such fluctuations are not incorporated into definition Eq.~\ref{logreturn}. However, in a first approximation, we can ignore these effects if the timescales are short enough. This nonlinear transformation yields other problems since it can change the statistical properties of the underlying process. Because randomness can affect the logarithmic growth rates of crime in a city in a size-dependent manner, a natural question arises: ``How city size affects the logarithmic returns of crime growth rates?''

\begin{figure}[!h]
\centering
\includegraphics[width=0.98\linewidth]{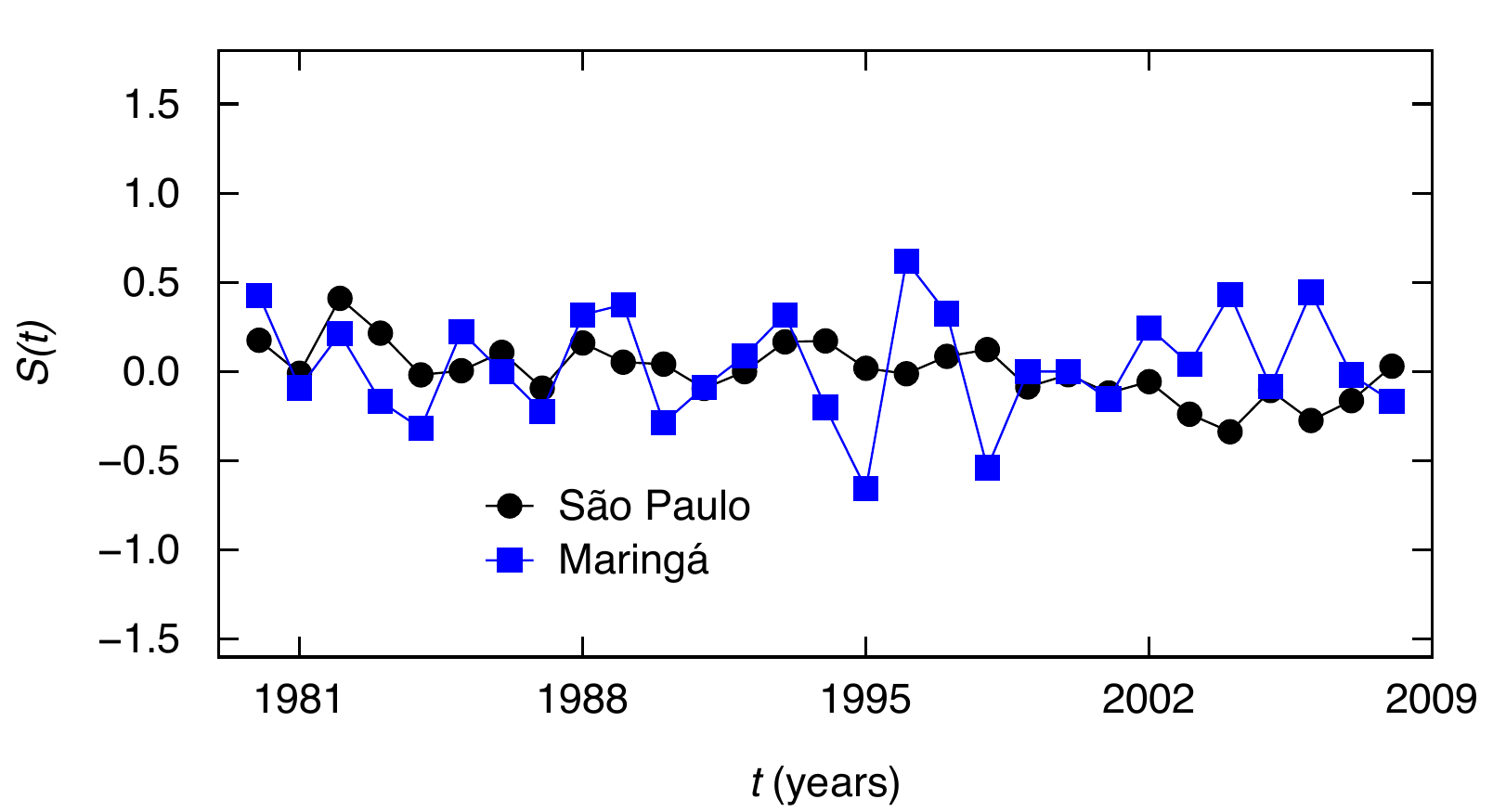}
\caption{Comparison between time series of homicides growth rates for two Brazilian cities. Mari\-ng\'a (blue line with square markers), with about 350 thousand inhabitants in 2010, shows much more variance than S\~ao Paulo (black line with circle markers), which in 2010 the population size were about 10.6 million people. Figure adapted from reference~\cite{Alves2013scaling}}
\label{fig:1}  
\end{figure}

To answer this question, let us first consider time series of homicides growth rates in cities with different populations size by using the definition of Eq.~\ref{logreturn}. Fig.~\ref{fig:1} shows a comparison between two cities with very distinct size, one with about 350 thousand inhabitants (Maring\'a) and another with about 10.6 million people (S\~ao Paulo), both values for the year 2010. The fluctuations in these time series are remarkably more prominent in Maring\'a than in S\~ao Paulo. Because city size seems to have an important role in the fluctuation, it is useful investigating them separately. Thus, we can group cities into $w$ categories (with $w=\{1,2,3,...,n\}$) that depends on the population size $N$. For example, in the category $w=n$ we only consider cities with size $N^*$ satisfying the relation $N_{w=n-1}<N^*<N_{w=n}$. We have omitted the time dependency of the population size because, in principle, we could use the population size for any $t$ in the range of the time series, although it is common to consider the initial value of the time series as an indicator of the organization size~\cite{Alves2013scaling,Stanley1998growth,Picoli2008universal,Picoli2006scaling,Labra2007scaling}.

Having the cities grouped into the $w$ categories, we calculate the standard deviation of the homicide growth rates, $\sigma (S)$, of cities with different population sizes $N$ to investigate how fluctuations affect the homicide growth rates. The distribution of crime growth rates is approximately described by a Laplace distribution $L(\mu,\sigma)$ (tent-shaped distribution) where the parameters $\mu$ and $\sigma$ are the average and standard deviation of the data~\cite{Alves2013scaling}. Moreover, the parameters of the Laplace distribution depend on the range of sizes $N^*$ selected. It turns out that this distribution can be rescaled by using the variable $s(t)=[S(t)-\mu]/\sigma$, an operation that collapses  all distributions of crime growth rates for different population size $N^*$ into a single curve. A similar scaling invariant behavior is also observed in other complex organizations such as firms~\cite{Stanley1998growth}, religion activities~\cite{Picoli2008universal}, paper citations~\cite{Picoli2006scaling}, and metabolic rates in biology~\cite{Labra2007scaling}.

\begin{figure}[!h]
\centering
\includegraphics[width=0.95\linewidth]{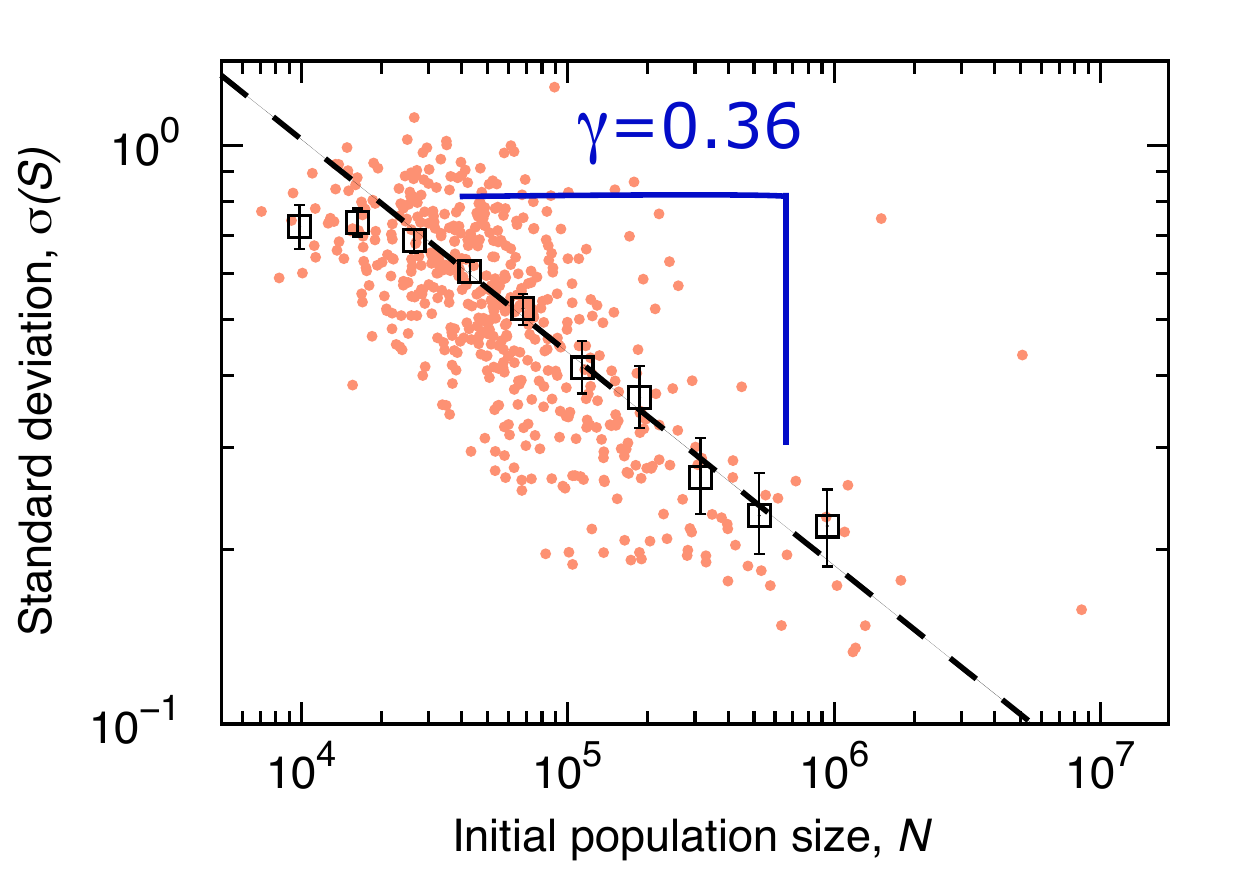}
\caption{Scaling behavior in the dynamics of crime growth rates. The relationship between the standard deviation of the logarithmic returns of crime is described by a power-law decay of the initial population size, this is, $\sigma(S(t)) \approx N^{-\gamma}$, with $\gamma = 0.36$. Figure adapted from reference~\cite{Alves2013scaling}}
\label{fig:2}  
\end{figure}

Another scaling property is obtained from the relationship between fluctuation in homicides growth rates and population size. Figure~\ref{fig:2} shows the relationship between the standard deviation of the logarithmic returns $\sigma[S(t)]$ and the population size $N$. The average trend of data is well described by a power-law function 
\begin{eqnarray}
\label{powerlawdecay}
\sigma \sim N^{-\gamma}
\end{eqnarray}
where $\gamma$ is the scaling exponent. 

The emergence of these scaling properties enable us to model crime rates in the framework of complex organizations with interacting subunits, similarly to physical systems where inanimate particles interacting to each other exhibit an emerging complex behavior~\cite{stanley1996scaling,amaral1997scaling,Stanley1998growth}. Usually, a murder case is preceded by a sequence of events, for instance, a discussion that became more aggressive, followed by a fight, and a murder, or a drug-dependent that not pay the drug-dealer and got murdered for revenge. There are indeed several other situations in which a sequence of events is followed by a homicide. Mathematically, this process can be described as a multiplicative process, where the probability of an event happens is dependent in a multiplicative fashion of the probabilities associated with a number of other events.

In 1931, Gibrat~\cite{Gibrat1931Les} proposed a model where the size of a complex organization depends on its previous size multiplied by Gaussian noise with zero mean and unitary variance, that is,
\begin{equation}
 S(t+\Delta t)=S(t)+A \epsilon (t) S(t)\,,
\end{equation}
where $S(t)$ is the organization size in time $t$, $A$ is a positive constant, and $\epsilon(t)$ is the Gaussian noise. This model assumes that the growth rates are completely uncorrelated in time, which is not true in the context of crime and in most complex organizations. A generalization of Gibrat's model that includes memory effects was proposed by Picoli \textit{et al.}~\cite{Picoli2006scaling} by using the following relation
\begin{equation}
\label{picolimodel}
 S(t+\Delta t)=S(t)+\lambda(t) S^k(t),
\end{equation}
where
\begin{eqnarray}
\lambda(t)=[A+B \lambda(t-1)]\epsilon(t),
\end{eqnarray}
$A$, $B$, and $k$ are positive constants, and $\epsilon(t)$ is a random number following a Gaussian distribution with zero mean and unitary variance. The limit where $k=1$ and $B=0$ recovers Gibrat's model. By comparing Eq.~\ref{picolimodel} with Eq.~\ref{powerlawdecay}, we note that $k=1-\gamma$ and thus we can re-write Picoli \textit{et al.} model as
\begin{equation}
 S(t+\Delta t)=S(t)+[A+B \lambda(t)(t-1)]\epsilon(t)] S^{1-\gamma}(t).
\end{equation}
This simple model reproduces key aspects of complex organizations growth such as the tent-shaped distribution of the growth rates and the power-law behavior of the standard deviation with organization size. 

\subsection{Scaling laws of urban metrics and crime with population size}
\label{subsec:1.2}

\begin{figure*}[!ht]
\centering
\includegraphics[width=0.9\linewidth]{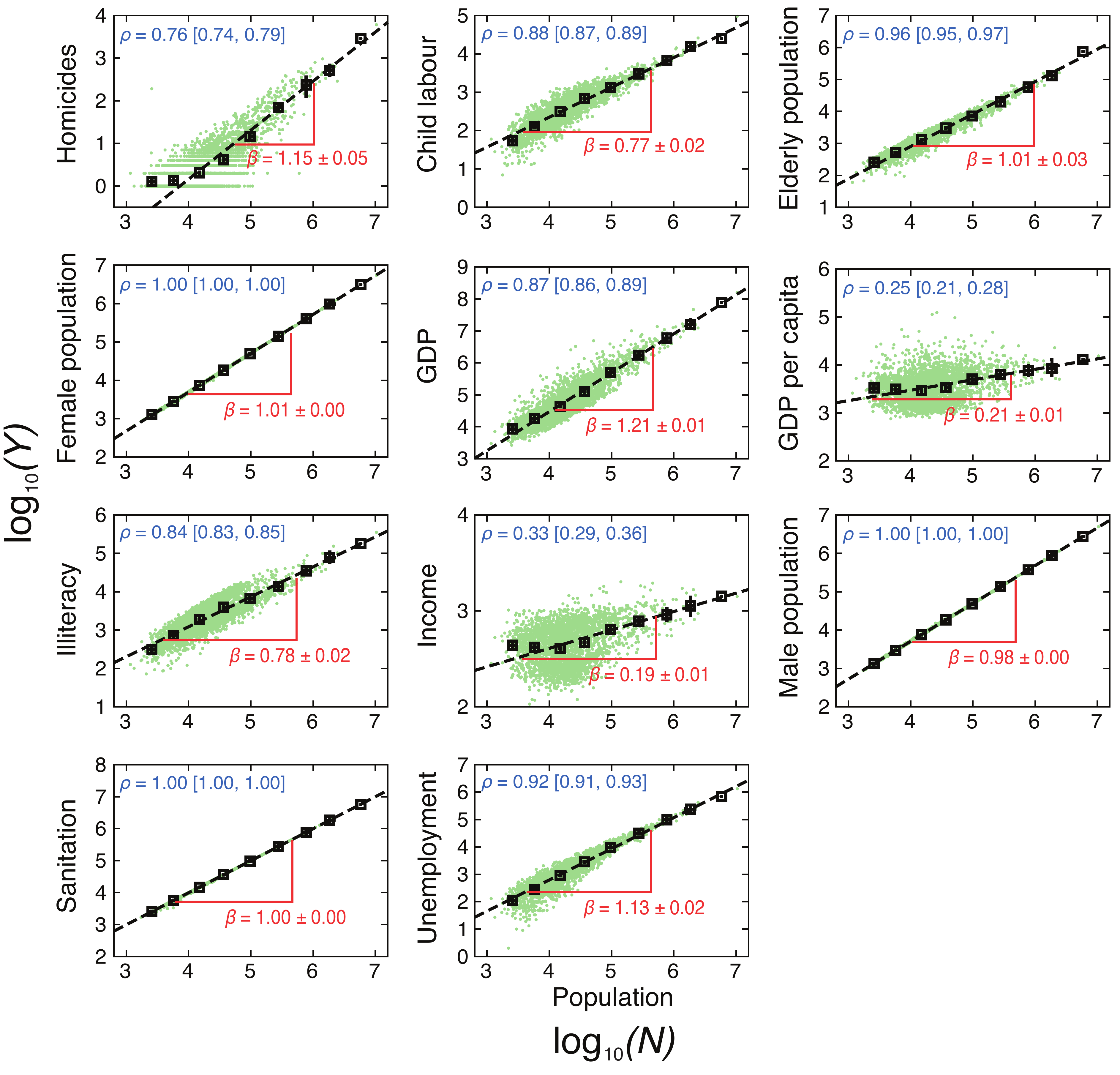}
\caption{Allometric scaling in the Brazilian cities. Figure reproduced from reference~\cite{Alves2013distance}}
\label{fig:3}  
\end{figure*}

Scaling laws in the relationships between urban indicators and city size are among the most interesting findings of recent studies on urban systems. Some examples of urban indicators that exhibit scaling laws with the population size includes patent, gasoline, gross domestic product~\cite{Bettencourt2007PNAS, Arbesman2009, Bettencourt2010nature, Bettencourt2010}, crime~\cite{Bettencourt2010, GomezLievano2012, Alves2013scaling,Alves2013distance,Alves2014empirical, Ignazzi2014,hanley2016rural}, educational indicators~\cite{Melo2014}, the number of candidates for the elections ~\cite{Mantovani2011, Mantovani2013}, transport networks ~\cite{Samaniego2008,Louf2014plosone}, employees from various sectors ~\cite{Pumain2006}, and social interaction measures~\cite{Pan2013}. Mathematically, these scaling laws between urban metrics $Y$ and population size $N$ are written as
\begin{equation}\label{eq:urbanscaling}
Y = \mathcal{A}\, N^{\beta}\,,
\end{equation}
where $\mathcal{A}$ is a constant and $\beta$ the scaling exponent. These scale invariant relationships summarize the average effects of population size on urban metrics. 

Scale invariance is an exact form of self-similarity, in which for every magnification (scale) there is a smaller part of the object which is similar to the whole. Self-similarity is a typical property of fractal geometries, in which parts of the object show similar properties on many scales~\cite{Mandelbrot1967}. In mathematics, an object is self-similar if it is exactly, or approximately, similar to a part of itself (that is, the whole has the same form as one or more parts). For example, we can consider the scaling properties of a function $f(x)$ under operations in the $x$ variable, that is, we want a function in the form $f(\lambda x)$ for some scale $\lambda$, where $\lambda$ can be a length, size, or energy. Usually, $f(x)$ is scale-invariant and self-similar if 
\begin{equation}
f(\lambda\, x) = \lambda^\delta f(x) \nonumber
\label{eq:homogenea}
\end{equation}
for some value of $\delta$ and $\lambda$, so that the above condition establishes a homogeneous equation of first order~\cite{callen1985thermodynamics}. 

Self-similar scaling laws (or allometric laws) have been found in several contexts, from biological~\cite{Kleiber1932, Kleiber1947, West1997, west2002allometric, west2005origin} to urban systems~\cite{Bettencourt2007PNAS, Bettencourt2010nature,Alves2013distance,Alves2013scaling}. One of the most famous examples of allometry was found by Kleiber in 1932~\cite{Kleiber1932} and is known as Kleiber's $3/4$ law. This allometric law states that the metabolic rate increases with the mammalian mass in a power-law fashion with exponent $3/4$. In other words, large animals are more efficient regarding energy consumption by body mass~\cite{Kleiber1932, Kleiber1947, West1997, west2002allometric, west2005origin}.

In the context of cities, similarly to Kleiber's law, large cities are more efficient (in per capita terms) regarding resource consumption such as the number of gasoline stations, length of power cables and road mesh area (sublinear relations)~\cite{Bettencourt2007PNAS}. On the other hand, large cities produce more wealth, patents, social and environmental problems (also per capita terms) than small towns (superlinear relations)~\cite{Bettencourt2007PNAS}. Figure~\ref{fig:3} shows examples of allometries in several urban indicators of Brazilian cities. In general, urban indicators can be classified into three categories: $\beta<1$ are indicators that exhibit an economy of scale and are usually associated with infrastructure, analogous to biological quantities; $\beta>1$ are indicators displaying increasing returns with population size such as GDP, crime, diseases, innovation or wealth, which are usually associated with the intrinsically social nature of cities; $\beta=1$ are indicators associated with individual human needs such as housing, jobs, or sanitation needs, as shown in Fig.~\ref{fig:3}.

\section{Modeling crime through urban metrics}
\label{sec:2}
Quantifying and predicting the performance of cities is a common problem addressed by researchers and governmental agencies. In the context of crime levels in cities, a question arises: ``What are the safest cities to live?.'' A typical answer to this question is found governmental crime reports, where cities are ranked according to per capita crime indicators or number of crimes per 100 thousand inhabitants. The problem in considering crime or any city indicator divided by population is that this approach explicitly assumes a linear relationship between the crime indicator and population size. It also implies that cities are extensive systems, that is, that their subunits behave as the whole system. However, the urban scaling hypothesis points out to the opposite direction: cities are non-extensive complex systems and their isolated parts do not behave in the same way as when they are interacting.

Another typical question in the context of modeling crime is ``Are wealthier cities safer?''. Indeed, this question can be generalized for considering any other city metric of interest, such as inequality, unemployment, and education metrics. Mathematically, we may seek for a function $f$ that returns the number of crimes $H$, given an urban metric, or, more generally, a set of urban metrics. Thus, considering $m$ urban metrics and approximating the $f$ function by a linear combination these urban metrics $Y$, we have the following linear regression model
\begin{eqnarray}\label{eq:regression}
H=f(Y_1,{\dots},Y_m)= C_0 + \sum_{i=1}^m C_i\, Y_i +\xi\,,
\end{eqnarray}
where $\xi$ is a random noise accounting for the unobserved determinants of crime, $Y_i$ is the $i$-th urban metric, $C_0$ is the intercept constant, and, usually, $C_i$ is associated with the predictive power that the indicator $i$ has to describe the number of crimes $H$. However, this model assumes again that urban metrics are independent of population size. In the following subsections, we describe an approach to overcome this problem and also present a model to quantify the influence of urban metrics on crime.

\subsection{Scaled-adjusted metric}
\label{subsec:2.1}
\begin{figure*}[!ht]
\centering
\includegraphics[width=0.9\linewidth]{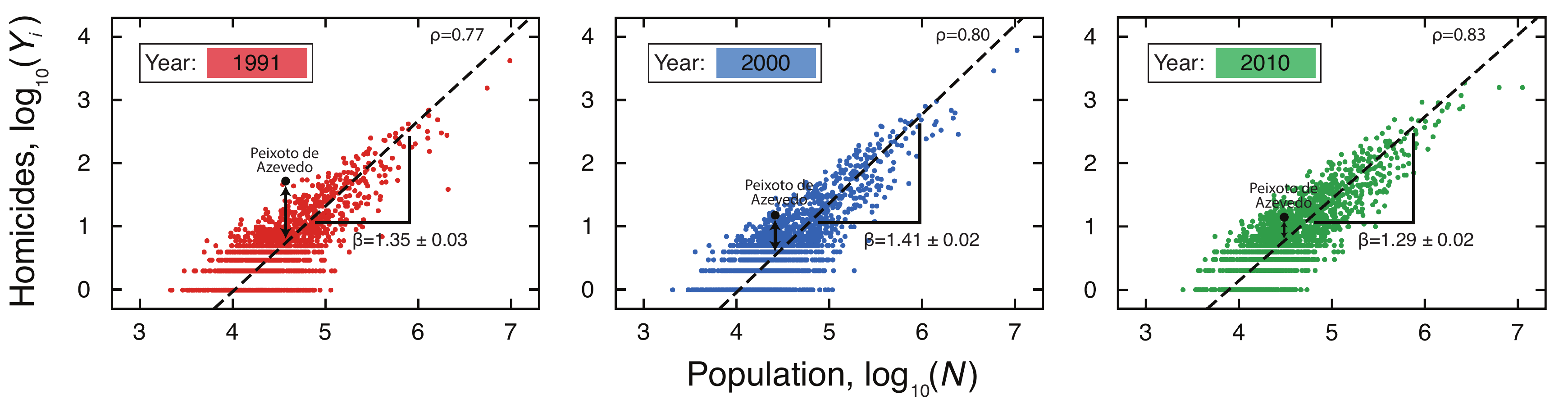}
\caption{Definition of scaled-adjusted metric for homicides in Brazilian cities in three different years $t$. The city \textit{Peixoto de Azevedo} is highlighted in these plots. Figure reproduced from reference~\cite{Alves2015scaledmetric}}
\label{fig:4}  
\end{figure*}

A simple and efficient way to overcome the non-linearities present in the relationships between urban metrics and population size is by considering the scale-adjusted urban metric~\cite{Alves2015scaledmetric,Alves2013distance,Bettencourt2010}. This quantity consists in evaluating the residuals of the adjusted scaling law in logarithmic scale, that is, the difference between observed empirical value of an urban metric $Y$ and the value expected by the allometric scaling with population size. Mathematically, we write
\begin{equation}
\label{eq:scaleadjusted}
D_Y(t)=\log Y(t)-\langle \, \log Y(t) \rangle
\end{equation}
where
\begin{equation}
\langle \, \log Y(t) \rangle = \log \mathcal{A}+ \beta \log N(t)
\end{equation}
represent the allometric scaling law. Fig.~\ref{fig:4} illustrates the definition of this scaled-adjusted metric.

Scaled-adjusted metrics are more suitable to linear regressions since it explicitly accounts for the effects of population size on urban metrics and crime. In contrast with the naive approach where no data transformation is used (usually yielding meaningless coefficients in the model of Eq.~\ref{eq:regression}), this approach not only provides better variables for performing linear regressions but also allows a fairer comparison of cities with different population size. By removing the population bias, the relationship of crime with urban metrics are now meaningful, and we can further interpret how a certain type of crime is related to a given urban metric. Also, scaled-adjusted metrics are linearly correlated with their past values, making them especially attractive to forecast crime indicators after a time interval $\Delta t$~\cite{Alves2015scaledmetric}, as we shall discuss later on. 

\subsection{Quantifying the influence of urban metrics on crime}
\label{subsec:2.2}

\begin{figure}[!h]
\centering
\includegraphics[width=0.8\linewidth]{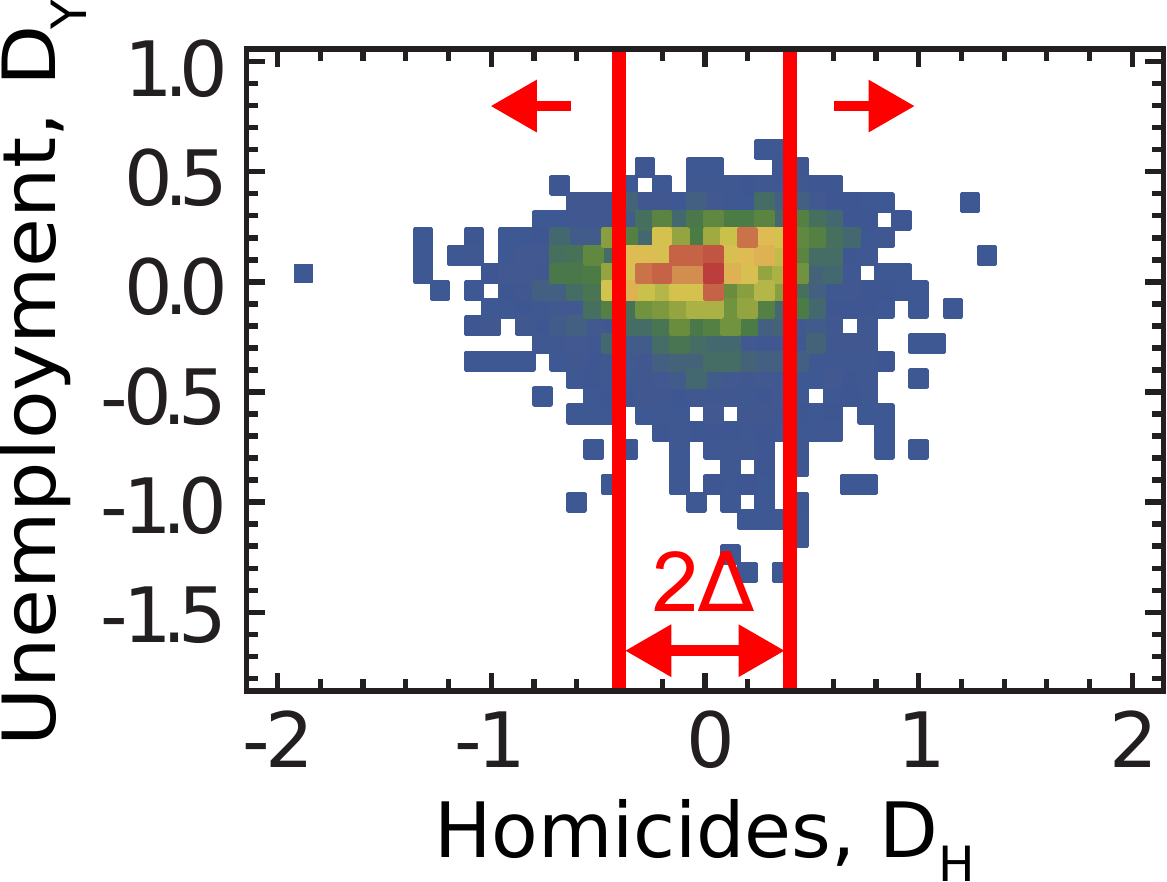}
\caption{Scaled-adjusted metrics for unemployment $D_Y$ versus the one for homicides, denoted as $D_H$. The red vertical lines indicate the threshold $\Delta$ for cities above (positive side) and below the scaling law (negative side) for the homicide indicator. Figure adapted from reference~\cite{Alves2013distance}}.
\label{fig:5}  
\end{figure}

Urban metrics and crime are described by scaling allometric laws, allowing us to calculate the scaled-adjusted metrics defined in Eq.~\ref{eq:scaleadjusted}. From the definition of scaled-adjusted metrics, we can categorize cities into two classes: Cities that are above the scaling law (that is, $D_Y>0$) and cities that are below the scaling law ($D_Y<0$). This procedure allows us to investigate how urban indicators are correlated with homicides after removing the effects of population size. Thus, we can verify, for example, whether cities that are going well in terms of GDP have more crime or whether cities that are going bad in terms of illiteracy rates have more crime. A straightforward approach to visualize the correlations  between the scaled-adjusted metric associated with crime and the ones from other urban metrics is by making a scatter plot, as depicted in Fig.~\ref{fig:5} for the case of unemployment.

The dispersion of data in Fig.~\ref{fig:5} hinders information about the relationship between crime and unemployment. To access whether the scaled-adjusted metrics of crime and unemployment are correlated in a significant manner, we consider a threshold of homicides $\Delta$ above and below the allometric power law (red lines in  Fig.~\ref{fig:5}). Mathematically, we seek for a function $f$ that describes the averages values of the scaled-adjusted metrics as a function of the homicides threshold, that is, $f=f(\Delta)$. Ignoring the signal (because we have grouped cities into above and below the scaling law), we can write this function as 
\begin{eqnarray}
f(\Delta)=E[ D_{Y}|D_H \geq \Delta]\,,
\end{eqnarray}
where $E$ is the expected value of the scaled-adjusted metric for a given urban metric $Y$ conditional to the $D_H \geq \Delta$ value. Thus, given a threshold for the variable related to homicides, we calculate the expected value of the scaled-adjusted metric for a given indicator for cities into the two classes, that is, for cities with a number of homicides above and below the allometry with population size.

\begin{figure*}[!ht]
\centering
\includegraphics[width=0.9\linewidth]{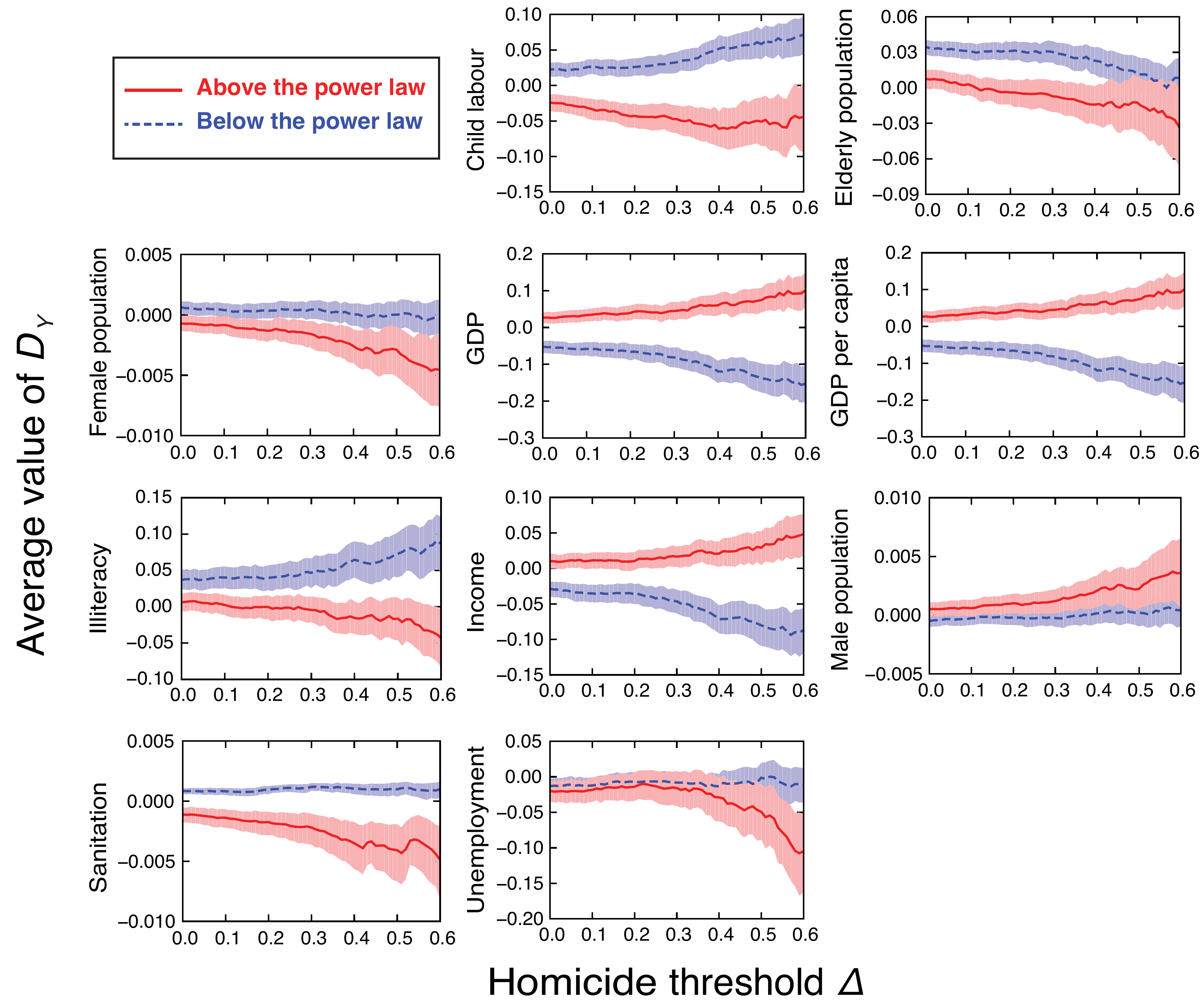}
\caption{The average values of the scaled-adjusted metric $D_Y$ evaluated for each urban indicator in function of the homicide threshold $\Delta$ after grouping the cities that are above (red continuous lines) and below (blue dashed lines) the scaling laws with the population size. Figure reproduced from reference~\cite{Alves2013distance}}
\label{fig:6}  
\end{figure*}
This approach is better to extract correlations among urban metrics and crime because population effects are removed by the scaled-adjusted metric defined in Eq.~\ref{eq:scaleadjusted}. By varying the homicide threshold according to the method depicted in Fig.~\ref{fig:5}, we obtain how the different urban indicators are related to crime. In Fig.~\ref{fig:6}, we show these relationships for several urban metrics obtained from Brazilian cities. This figure indicates that there are three types of correlations between crime and urban metrics. In the first one, when the homicide threshold increases, the scaled-adjusted metrics of cities above the scaling laws also increase, whereas the values of scaled-adjusted metrics for cities below the scaling laws decrease. That is the case of GDP, GDP per capita, income, and male population, which indicates that these metrics have positive correlations with crime. In the second one, when the homicide threshold increases, the scaled-adjusted metrics of cities above the scaling laws decrease, whereas the values of scaled-adjusted metrics for cities below the scaling laws increase. That is the case of child labor, female population, illiteracy, and sanitation indicators; and indicates a negative correlation with crime. The last type of correlation between urban metric and crime is the case of elderly population and unemployment, which have only significant power to describe crime in a certain range of the homicide scaled-adjusted metric. Specifically, the correlation between unemployment and crime is only significant when the homicide threshold exceeds $\Delta \approx 0.56$, and elderly population does not present significant correlation above $\Delta \approx 0.45$.

\subsection{Predicting crime through urban metrics}
\label{subsec:2.3}

\begin{table*}[!ht]
\renewcommand{\arraystretch}{.8}
\caption{{\bf Regression model coefficients [$D_{Y_i}(2010)$ versus $D_{Y_i}(2000)$] for the indicator homicides.} Values of the linear coefficients $C_k$ obtained via ordinary least-squares fit and their  standard errors. Here, $z$ is the value of the $z-$statistic and $p$ is the two-tail $p-$value for testing the hypothesis that the coefficient $C_k$ is different from zero.}
\centering
\begin{tabular}{lrrrr}
\hline
Scaled-Adjusted Urban Metric & Coefficient $C_k$ & Standard Error  & z-Statistic & $p>|z|$\\\hline
Intercept, $C_0$ & $<10^{-4}$&0.0069&-0.0017& 0.9987\\
Child labor & 0.0682&0.0449&1.5195& 0.1286\\
Elderly population & -0.862&0.0958&-8.994& $<10^{-4}$\\
Female population & 6.5982&14.0685&0.469& 0.6391\\
Homicides & 0.3028&0.0189&15.9874& $<10^{-4}$\\
Illiteracy & 0.5673&0.0478&11.8719& $<10^{-4}$\\
Family income & 0.1361&0.0556&2.4497& 0.0143\\
Male population & 3.337&14.3795&0.2321& 0.8165\\
Unemployment & 0.189&0.0417&4.5345& $<10^{-4}$\\
\hline
\multicolumn{5}{r}{Adjusted $R^2 =0.395$ }\\
\hline
\end{tabular}
\label{tab:s2010vs2000_4}
\end{table*}

Having defined an appropriate metric for describing the relationships among urban metrics and crime, we can reformulate the model proposed in Eq.~\ref{eq:regression} in terms of the scaled-adjusted metrics. Basically, we replace the $Y$ variables in Eq.~\ref{eq:regression} by the correspondent scaled-adjusted metric $D_Y$, and the response variable becomes the scaled-adjusted metric for homicides $D_H$ instead of the raw number of homicides $H$. Because we aim to predict future values of crime indicators, we introduce a time interval $\Delta t$ in the response variable of crime. Thus, taking into account the population size effects on urban metrics and crime, we write the following model
\begin{equation}\label{eq:scaledregression}
D_{H}(t+\Delta t)= C_0(t) + \sum_{i=1}^N C_i(t)\, D_{Y_i}(t) +\eta,
\end{equation}
where $C_0$ is the intercept, $C_i$ is the regression coefficient for each scaled-adjusted metric $D_{Y_i}$, $\eta$ is a random noise, and $t$ stands for time.

The model of Eq.~\ref{eq:scaledregression} was proposed in~\cite{Alves2015scaledmetric} to reproduce several patterns of the data associated with Brazilian cities. For instance, by using this approach, the authors showed that it is possible to predict the average values of homicides with great precision when grouping cities in above and below the scaling law, and that the model reproduces approximately the distribution of $D_H(t+\Delta t)$. In Table~\ref{tab:s2010vs2000_4}, we reproduce the values of the linear coefficients $C_i$ obtained via ordinary least-squares fitting the model from Eq.~\ref{eq:scaledregression} for $t=2000$ and $\Delta t=10$ years.

\begin{figure}[!h]
\begin{center}
\includegraphics[width=0.95\linewidth]{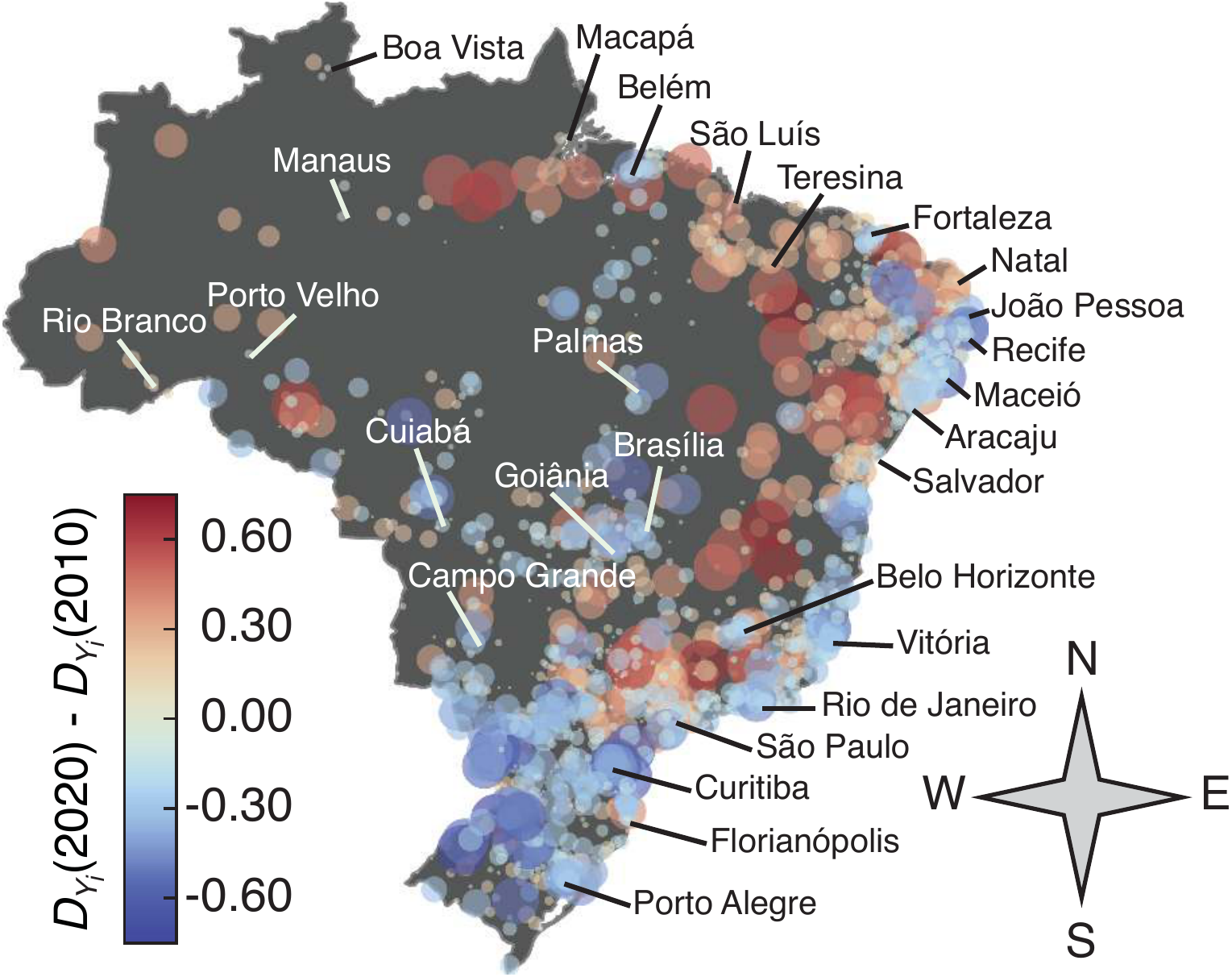}
\end{center}
\caption{Spatial visualization of the differences between the forecast scaled-adjusted metrics for homicides and the empirical ones, that is, $D_Y(2020)-D_Y(2010)$. The size of the circles represent the change magnitude and colors represent the signals (blue shades indicate negative growth and red represent positive growth). Figure adapted from reference~\cite{Alves2015scaledmetric}}
\label{fig:7}  
\end{figure}

It is worth noting that there is an strong correlation between the scaled-adjusted metric of homicides in the year $t+\Delta t$ and the year $t$. Elderly population, illiteracy, family income, and unemployment are also important for describing crime. However, we further observe that in spite of scaled-adjusted metrics removing the population size effects, this approach does not eliminate multicollinearity in the data, and further statistical procedures are required in order to have a correct interpretation of the importance of urban metrics to describe crime~\cite{Alves2017statistical}.

It is striking that predicted changes appear spatially clustered, despite the absence of spatial variables in the model, a result that indicates the existence of spatial correlations and collective dynamics~\cite{Alves2015spatial}. The model predicts a decrease in $D_H$ for the vast majority of southern cities, and densely populated cities near the coast, from Rio de Janeiro to Jo\~ao Pessoa. Inner cities, especially the ones from S\~ao Paulo State and Northeastern Region, are expected to increase $D_H$, suggesting that this violent crime is ``moving" towards less populated areas of the interior of Brazil~\cite{Alves2015scaledmetric}.

\section{Perspectives on crime modeling through the lenses of complex systems}
\label{sec:3}

Understanding and preventing crime remains a major challenge for society, and despite recent advances obtained in criminology about the mechanisms of crime, we still need more empirical investigations and model validation for a better understanding of crime and its relationships with socioeconomic indicators. We believe that with the increasing amount of data related to crime, the use of tools from statistical physics, complex systems, and network science is not only suitable but necessary for understanding, modeling, and preventing crime. These tools allow us to explore crime at different levels of society, from countries and cities to criminals, and also modeling individual criminal patterns to establish laws describing the emergent behavior of crime interactions in the different levels of society. Such investigations can have a direct impact on how we allocate security resources and how we make rules for preventing crime in cities. 

\textbf{Acknowledgments:} L.G.A.A. acknowledges FAPESP (Grant No. 2016/16987-7) for financial support. H.V.R. acknowledges CNPq (Grant No.  440650/2014-3) for financial support. F.A.R. acknowledges CNPq (Grant No.  307748/2016-2) and FAPESP (Grant No. 2016/25682-5 and Grant No. 13/07375-0) for financial support.

\bibliography{ref}

\end{document}